# Creation of independently controllable and long lifetime polar skyrmion textures in ferroelectric-metallic heterostructures

*Fei Sun, Jianhua Ren, Hongfang Li, Yiwei Wu, Jianwei Liang, Hui Yang, Yi Zhang, Jianyi Liu, Linjie Liu, Mengjun Wu, Xiaoyue Zhang, Wenpeng Zhu, Weijin Chen,* * *and Yue Zheng* *

F. Sun, J. Ren, Y. Wu, J. Liang, H. Yang, Y. Zhang, L. Liu, M. Wu, X. Zhang, W. Zhu, W. Chen, Y. Zheng

Guangdong Provincial Key Laboratory of Magnetoelectric Physics and Devices, State Key Laboratory of Optoelectronic Materials and Technologies, Centre for Physical Mechanics and Biophysics, School of Physics, Sun Yat-sen University, Guangzhou 510275, China

E-mail: chenweijin@mail.sysu.edu.cn and zhengy35@mail.sysu.edu.cn

H. Li

School of Materials Science and Engineering, Chongqing Jiaotong University, Chongqing 400074, China

J. Liu

College of Physics, Qingdao University, Qingdao 266071, China

M. Wu, W. Chen

School of Materials, Shenzhen Campus of Sun Yat-sen University, Shenzhen 518107, China

**Abstract**

Topological textures like vortices, labyrinths and skyrmions formed in ferroic materials have attracted extensive interests during the past decade for their fundamental physics, intriguing topology, and technological prospects. So far, polar skyrmions remain scarce in ferroelectrics as they require a delicate balance between various dipolar interactions. Here, we report that $PbTiO_3$ thin films in a metallic contact undergo a topological phase transition and stabilize a broad family of skyrmion-like textures (e.g., skyrmion bubbles, multiple π-twist target skyrmions, and skyrmion bags) with independent controllability, analogous to those reported in magnetic systems. Weakly-interacted skyrmion arrays with a density over 300 Gb/inch$^2$ are successfully written, erased and read-out by local electrical and mechanical stimuli of a scanning probe. Interestingly, in contrast to the relatively short lifetime <20 hours of the skyrmion bubbles, the multiple π-twist target skyrmions and skyrmion bags show topology-enhanced stability with lifetime over two weeks. Experimental and theoretical analysis implies the heterostructures carry electric Dzyaloshinskii-Moriya interaction mediated by oxygen octahedral tiltings. Our results demonstrate ferroelectric-metallic heterostructures as fertile playground for topological states and emergent phenomena.

## 1. Introduction

Ferroelectrics are considered as promising alternatives of magnetic systems to stabilize topological textures such as vortices, labyrinths and skyrmions. During the past decade, they have spurred tremendous research interests for searching exotic topological polar textures and emergent phenomena, as well as potential device applications.[1,2] Among the rich topological textures, polar skyrmions are special swirling polarization textures with particle-like topology, and they are shown to have a series of intriguing properties such as nanometric size,[3,4] switchable polarity and vorticity,[5,6] negative permittivity,[7] configurable electromechanical responses,[8] etc., holding great promise in developing topologically robust devices. However, polar skyrmions remain scarce in ferroelectrics. So far, they have been mainly found in ferroelectric-insulator heterostructures (FI) like $PbTiO_3/SrTiO_3$ (PTO/STO) superlattices and multilayers.[3,4,9] In such systems, polar skyrmions are collectively stabilized in a dense state to balance the elastic, electrostatic and gradient energies of the system. This is different from their magnetic counterparts, which are driven by the Dzyalozhinskii-Moriya interaction (DMI).[10-14] Consequently, polar skyrmions in superlattices and multilayers typically show strong interactions with their neighbors, making it challenging to independently manipulate the state of a single polar skyrmion by local electrical/mechanical stimuli without affecting its neighbors,[15-18] limiting their application prospects in storage and logic devices. The use of insulating layers also limits applications based on the electronic characteristics of polar skyrmions.

A fundamental issue of polar skyrmions is therefore: if there are any other systems beyond FI heterostructures can hold polar skyrmions, especially with weak interactions between the skyrmions like their magnetic counterpart? Recent studies have suggested that all-perovskite ferroelectric-metallic (FM) heterostructures such as $PbTiO_3/SrRuO_3$ (PTO/SRO) and $BaTiO_3/SrRuO_3$ (BTO/SRO), are promising candidates to carry exotic topological states and emergent phenomena.[19,20] For example, skyrmion-like bubbles and Labyrinth-like patterns were observed in FM heterostructures, and the presence of magnetic and electric DMI (eDMI) mediated by oxygen octahedral tiltings (OOT) has also been recently predicted by symmetry analysis and *ab initio* calculation.[21-24] Here, combining piezoresponse force microscopy (PFM), transmission electron microscope (TEM), density functional theory (DFT) calculations and phase-field (PF) simulations, we show that PTO thin films on top of metallic $Ca_{0.94}Ce_{0.04}MnO_3/SrRuO_3$ (CCMO/SRO) layer exhibit a delicate topological phase transition and can stabilize a broad family of skyrmion-like textures with independent controllability, including Néel-type skyrmion bubbles (with topological charges $Q = \pm1$), multiple π-twist target skyrmions (2π, 3π, 4π, …), and skyrmion bags, analogous to those discovered in magnetic systems.[25-28] Desirable feature of the present system for storage and logic applications are that one can feasibly write and erase skyrmions with regular distribution, weak coupling, and high density (over 300 Gb/inch$^2$) by local electrical or mechanical stimuli of a scanning probe microscope (SPM) tip, and that the skyrmion bits can be read out by their high conductance compared with the background. Interestingly, the stability of the skyrmion textures is found to be correlated with their topological degree. Contrast to the relatively short lifetime (< 20 hours) of the skyrmion bubbles, multiple π-twist target skyrmions and skyrmion bags exhibit much longer lifetimes (> two weeks), indicating a topology strategy to encode robust information in ferroelectrics.

## 2. Results and Discussion

### 2.1. Topological phase transition driven by electrical and mechanical stimuli

PTO/CCMO/SRO heterostructures were epitaxially grown on single-crystal (001) STO substrates via pulsed laser deposition (PLD) method (see Experimental Section; and Figure S1a, Supporting Information). Thickness of SRO layer was controlled to be ~20 nm and that of CCMO layer is less than 0.8 nm. The high-quality epitaxial growth of the heterostructures with atomically flat surface and interface was confirmed by atomic force microscopy (AFM) topography images (**Figure 1**a), the θ-2θ symmetric scan X-ray diffraction (XRD) patterns (Figure 1b), and high-angle annular dark-field scanning transmission electron microscopy (HAADF-STEM) images (Figure 1e). The PTO film exhibits a good ferroelectric property with coercive voltages of -2 and 3 V as shown by the piezoresponse force microscopy (PFM) phase hysteresis and amplitude butterfly loops (Figure 1d). Both the vertical PFM (VPFM) and lateral PFM (LPFM) phase images of the PTO film show uniform contrast (Figure 1c), and the HAADF-STEM images show that the polarization vectors of all the PTO unit cells point downward with small in-plane tilting (Figure 1e), together indicating that the pristine PTO thin film forms a homogeneous downward polarization state. The thin CCMO layer has a great impact on the polarization stability of the heterostructures. To see this, PTO/SRO heterostructures without CCMO layer were grown under similar conditions. The PTO/SRO heterostructures exhibit a homogeneous upward polarization state, which is opposite to the PTO/CCMO/SRO heterostructures (Figure S1b, Supporting Information). This reflects that the interfacial environments (both electrostatic and chemical) are modified by the inserted CCMO layer, which further changes the orientation of the out-of-plane polarization of the PTO thin film.[29-31] As the CCMO layer is only 1-2 unitcells thick, the PTO film is expected to be at a subtle balance between two polarization states with a moderate built-in field. The interfacial effect of CCMO layer is verified by DFT calculation (Figure S1c, Supporting Information). Additionally, DFT calculation also indicates that the PTO film probably possesses an interlayer electric DMI (eDMI) mediated by $a^-a^+a^-$ mode OOT (Figure S2 and Table S1, Supporting Information). Integrated differential phase contrast STEM (iDPC-STEM) confirms that the PTO thin film possesses large distortion of the oxygen octahedra (Figure 1f). We believe these effects collectively result in the PTO thin film being in a delicate state prone to topological phase transition.

We performed *in situ* TEM experiment to gain a first insight into the field-induced topological phase transition in the PTO thin film. Similar to the as-grown film, the TEM sample shows a uniform downward polarization state. A tungsten probe applies ± 3.7 V out-of-plane pulsed bias to the surface of the TEM sample to see if topological textures can be electrically induced. Evolution of the TEM contrast of the sample during the experiment is recorded in Movie S1 (Supporting Information), and typical snapshots of the process are displayed in Figure 1g. It shows that the TEM sample undergoes repetitive polarization switching under the alternating pulsed bias. The light and darkness contrasts of the TEM images indicate different domains, and the switched domain pattern probably consists of four different polarization directions (see schematic diagrams in Figure 1h), suggesting a center-divergent topological texture. Interestingly, accompanying with the repetitive creation and annihilation of the topological texture, the PTO film undergoes repetitive conductance switching (Figure S3a, Supporting Information). The topological texture therefore has a high conductance compared to the downward polarization background. Interestingly, the current evolution curve shows oscillating feature during the application of negative bias (Figure S3a, Supporting Information), indicating that the created topological texture oscillates continuously. This contrasts with the conventional 180° domain switching in PTO/SRO system (Movie S2 and Figure S3b, Supporting Information).

The formation of topological texture revealed by our *in-situ* TEM experiment is consistent with PFM experiment. As shown in the out-of-plane (left) and in-plane (right) PFM images in Figure 1i, the pristine downward polarization state (yellow-colored area) of the PTO film can be fully switched into upward polarization state (purple-colored area within the gray dashed box) under the scanning electric field (-3 V) applied by the PFM conductive tip. The upward polarization state remains stable under subsequent application of ± 2 V bias (the green dashed-line area). Interestingly, accompanied with the above electrical stimulation, two types of skyrmion-like bubbles with opposite polarity are clearly observed, in either the upward polarization domain (highlighted by red boxes) or downward polarization domain (highlighted by blue box) as shown in Figure 1i. To see them more clearly, zoom-in PFM phase and amplitude images of the two types of skyrmion-like bubbles are shown in **Figure 2**a, respectively. For the skyrmion-like bubbles with an upward polarity (highlighted by blue box), its LPFM phase image shows a half-violet and half-yellow contrast, and the LPFM amplitude image shows a round-shaped bright contrast splitted by a dark line, indicating a phase inversion of the lateral polarization component along the direction perpendicular to the PFM cantilever. Moreover, the corresponding VPFM phase image shows a 180° phase inversion from the center of the texture to the edge region.

Angle-resolved PFM (vector PFM) is further performed by rotating the PFM cantilever with a set of angles relative to the sample (Figure S4, Supporting Information).[32,33] It is found that the LPFM phase images always show two semicircles with opposite contrasts and their boundary (dark line) rotates with the cantilever. Similar PFM signals are found for skyrmion-like bubbles with a downward polarity (highlighted by red box), but with opposite phase contrast. These features are consistent with the simulated PFM signs of Néel-type skyrmion textures obtained by phase-field model as shown in Figure 2b (see Figure S5 for more details of the phase-field simulation, Supporting Information), and are analogous to those observed in previous studies.[3,4] Following the definition of Nahas et al.,[34] the skyrmion-like bubbles with an upward polarity at the center have a topological charge $Q = +1$, and those with a downward polarity have a topological charge $Q = -1$. It is noteworthy that a significant number of skyrmion-like bubbles also emerge at the domain walls between the upward and the downward polarization domains (see the blue arrows in Figure 1i) and some of them can further evolve into elongated skyrmion-like stripes (highlighted by pink/green boxes). The zoom-in PFM phase and amplitude images of the elongated skyrmion-like stripes and the phase field simulated PFM signals are also shown in Figure 2a,b, respectively. The schematics of the skyrmion-like textures are shown in Figure 2c. These results clearly show that the PTO thin film undergoes a delicate topological phase transition, which can breed rich topological defects.

Note that the topological phase transition of the PTO thin film can be also driven by mechanical stimulus, with the emergence of skyrmion-like bubbles and elongated skyrmion-like stripes similar to those induced by electrical stimulus. As shown in Figure 1j,k, a region of the PTO thin film is first poled into upward polarization state (the purple-colored region in the VPFM phase image), and then the film is scanned by the tip with different loading forces in a 1 μm × 1.5 μm area (white dashed box). It is found that the film occurs a large-area mechanical switching from upward polarization state to downward polarization state when the tip force reaches 2100 nN. Such a mechanical switching can be attributed to the flexoelectric effect.[35] Interestingly, topological phase transition is also triggered by the mechanical stimulation. After scanning by a tip force of 1000 nN, a skyrmion-like bubble with $Q = -1$ appears in the upward polarization region, meanwhile dense skyrmion-like bubbles with $Q = +1$ emerge in the downward polarization region. When the tip force increases to 1500 nN, elongated skyrmion-like stripes begin to form in the upward polarization region. As the tip force further increases to 2100 nN, the film undergoes large-area mechanical switching. During this switching process, skyrmion-like bubbles with $Q = +1$ as well as elongated skyrmion-like stripes are induced. Note that our results seem similar but actually not the same with the results reported by Zhang et al.,[36] who recently showed that applying AFM tip force can induce the formation of skyrmion-like bubbles with $Q = -1$ in PTO/SRO heterostructures with an initial upward polarization state. Firstly, we found that skyrmion-like bubbles with $Q = +1$ can also be created by applying AFM tip force. Secondly, skyrmion-like bubbles with $Q = +1$ are found to be more easily created by the tip force than their counterparts with $Q = -1$. This indicates that the tip-force-induced flexoelectric field is not the reason for the creation of skyrmion-like bubbles.

**2.2 Construction of a broad family of skyrmion-like textures**

In addition to the stabilization of skyrmion-like bubbles with topological charges $Q = \pm 1$, a broad family of skyrmion-like polar textures can be constructed in the PTO thin film by local electrical and mechanical stimuli exerted by PFM tip as shown in Figure 2d; and Figure S6, Supporting Information. Within the downward domain background, one can write textures include the simplest Néel-type skyrmion bubbles with $Q = +1$ (denoted as $\pi$-SK$^+$ for short, and superscript "+" indicates an upward polarization at the core), the multiple $\pi$-twist target skyrmions with $Q = 0$ (denoted as $2\pi$-TS$^-$, $4\pi$-TS$^-$ and $6\pi$-TS$^-$, with a downward core polarization and multiple-$\pi$ twist of the out-of-plane polarization from the core to the edge), as well as skyrmion bags with $Q = n - 1$ (denoted as SK$^+$ bags, and $n$ is the number of $\pi$-SK$^+$ bubbles enclosed by the ring). Moreover, skyrmion-like textures (e.g., $\pi$-SK$^-$, $3\pi$-TS$^-$, $4\pi$-TS$^+$, $6\pi$-TS$^+$, and SK$^-$ bags) can be also created in an electrically poled upward domain background. The LPFM phase images of these high-degree skyrmion-like textures are shown in Figure 2d, and their VPFM phase images can be found in Figure S6b (Supporting Information). The schematics of these skyrmion textures are also depicted at the bottom panels of Figure 2d. These skyrmion-like polar textures are analogous to those reported in magnetic materials, indicating the similarity between the two kinds of systems and the PTO/CCMO/SRO heterostructures are fertile playground for polar topological states.

**2.3 Regular patterning, control, and read-out of individual polar skyrmions**

Regular patterning, control (writing and erasure), and read-out of individual polar skyrmions are all necessary for developing skyrmion-based logic and memory devices. Nevertheless, it is challenging to realize these operations in superlattice and multilayer systems, where polar skyrmions collectively emerge with strong interactions. Here, we show that individual skyrmion-like bubbles can be precisely created in the PTO thin film, forming regularly distributed and weakly interacted arrays. In **Figure 3**a, skyrmion-like bubbles are precisely created to form a word "SKYRMION". Here, each capital letter consists of a few individual skyrmions, which are written by single-point loading of −2 V tip bias (detailed PFM results can be found in Figure S4, Supporting Information). Other regular skyrmion patterns, such as word "CPMB" (abbreviation of our laboratory name) and word "SYSU" (abbreviation of our university name) were also successfully written (Figure S7, Supporting Information). Remarkably, we can pattern the skyrmion-like bubbles with a density ~300 Gbit/inch$^2$ (Figure 3b), indicating that the real size of the written skyrmions is smaller than 50 nm and they have great potential in developing high-density memory devices. Moreover, besides the regular patterning of skyrmions, one can locally erase and read the skyrmion bits. As shown by the in-plane PFM amplitude and phase images in Figure 3c, −2 V tip bias was firstly applied to the PTO thin film at specific positions to create skyrmions, forming letter "R". After applying single-point tip bias (+2V) to the skyrmions within the yellow box, letter "R" was then changed to letter "I". Then, single-point tip bias (−2 V) was applied to rewrite two skyrmions, thus changing letter "R" to letter "r". Finally, consistent with the *in-situ* TEM experimental result (Figure 1g), CAFM measurement shows that the negative-bias induced skyrmion-like bubbles have a much higher conductance than that of the downward domain background (Figure S8, Supporting Information). Such a polarization-dependent transport behavior can be explained by the change of the Schottky barrier heights at the heterojunction interfaces. One therefore can achieve nondestructive read-out of the skyrmion bits based on the current at small bias.

**2.4 Lifetime of the skyrmion-like textures and topology-enhanced stability**

A remaining issue is the lifetime of the constructed metastable skyrmion-like polar textures, which is crucial for the design of skyrmion-based devices. To gain a complete picture of the stability of skyrmion-like polar textures in the PTO thin film, the temporal evolution of a series of topological textures is investigated and summarized. In **Figure 4**a, the temporal evolution of LPFM amplitude images of ten typical topological textures (including π-$SK^+$, 2π-$TS^-$, 4π-$TS^-$, 6π-$TS^-$, $SK^+$ bags, rings, etc.) written in a downward domain background are depicted. For simplicity, we use Roman numerals to label these textures. Specifically, **i** corresponds to Néel-type skyrmion-like bubbles (π-$SK^+$), **ii** to **vi** correspond to high-degree twisted skyrmions (2π-$TS^-$, 4π-$TS^-$, 6π-$TS^-$, 8π-$TS^-$ and 10π-$TS^-$), **vii** and **ix** correspond to a closed ring and a ring with a notch (unclosed ring), **x** corresponds to a 10π-$TS^-$ texture with unclosed rings, **viii** corresponds to a skyrmion bag with four π-$SK^+$ inside (there are also four π-$SK^+$ outside the ring for comparison).

One can see that the Néel-type skyrmion-like bubbles (**i**) begin to shrink after ~ 10 hours and ultimately annihilate after 20 hours. The 2π-$TS^-$ skyrmions (**ii**) show a slower relaxation process. They gradually relax into skyrmion-like bubbles (π-$SK^+$) after 20 hours and completely disappear after 60 hours, implying a higher stability brought by the higher topological degree. For the 4π-$TS^-$, 6π-$TS^-$, 8π-$TS^-$ and 10π-$TS^-$ skyrmions (**iii** to **vi**), their PFM images maintain almost unchanged after 400 hours (i.e., more than two weeks), revealing that the stability of skyrmions can be significantly enhanced by increasing the layers of π-twists (the number of outer rings). Moreover, the skyrmion bags (**viii**) also show lifetimes longer than two weeks. One can see that the stability of π-$SK^+$ skyrmion bubbles inside the ring is significantly enhanced (**viii**), in contrast to those outside the ring, which shrink and annihilate within one day. To confirm this effect, Figure 4b depicts another skyrmion bag texture, which has one “SYSU” pattern of π-$SK^+$ inside the ring and one outside the ring.

Similarly, the "SYSU" pattern outside the ring gradually annihilates within 60 hours, meanwhile the "SYSU" pattern inside the ring remains stable for more than two weeks. The ring structure hence provides a topology protection mechanism which slows down the collapse of the metastable skyrmion-like textures.

Two possible effects can be related to such a topology protection, as schematically illustrated in Figure 4c,d, respectively. On the one hand, there should be a topological protective barrier between the closed ring and unclosed ring structures (Figure 4c). The unclosed ring is topologically equivalent to elongated skyrmion stripe or bubble, which have shorter lifetimes. If we cut the closed ring with a notch by applying a local tip bias, its topology protection is broken. The unclosed ring eventually splits into short-lifetime elongated skyrmion stripes or skyrmion bubbles, as shown in Figure 4a (**vii**) and (**ix**). The poor stability of the unclosed ring leads to a relatively short lifetime of the $10\pi$-$TS^-$ textures with unclosed rings (**x**), which were also found to relax into $\pi$-$SK^+$ skyrmion bubbles and disappear within 60 hours. On the one hand, due to the electrostriction effect, the in-plane polarization of the ring structure compresses the region inside the ring (Figure 4d), leading to larger coercive fields and enhanced out-of-plane polarization inside the ring. Note that a larger tip bias of -3 V was required to create $\pi$-$SK^+$ bubbles inside a ring to form a skyrmion bag, compared with the tip bias of -2 V for the creation of $\pi$-$SK^+$ bubbles outside the ring. This reflects the coercive fields of the PTO thin film are indeed modulated by the ring structure. To see it clearly, we performed a spatial mapping of the local coercive voltages inside a series of ring structures with different sizes. The coercive voltages are extracted from the local PFM hysteresis loops collected at 20×20 locations. Statistic distributions of the coercive voltages are shown in Figure 4e (see detailed results of the PFM measurement in Figure S9,S10, Supporting Information). One can see that, as the radius of the skyrmion ring increases, the average coercive fields inside the ring significantly decrease, especially for that of up-to-down switching. This increase of the coercive field inside the ring, together with the intrinsic topology protection of the ring, leads to the topology-enhanced stability of the multiple $\pi$-twist target skyrmions and skyrmion bags. Therefore, one can use a topology strategy to construct long-lifetime skyrmion textures.

We also investigate the stability of skyrmion textures embedded in an electrically poled upward domain background. It shows that these skyrmion textures generally have a much weaker stability than the textures embedded in a downward domain background (Figure S11, Supporting Information). This difference likely originates from the intrinsic asymmetry of the two polarization directions of the PTO/CCMO/SRO heterostructure which favors a downward polarization state. The upward polarization state is less stable and is easier to trigger topological phase transition. Nevertheless, this also leads to a weaker stability of the artificially created skyrmion textures.

## 3. Conclusion

Combining SPM experiment, *in situ* TEM experiment, DFT calculation and phase-field simulation, we discover that PTO/CCMO/SRO heterostructures exhibit a delicate topological phase transition and can hold a broad family of skyrmion-like polar textures, including skyrmion bubbles ($Q$ =±1), high-degree twisted skyrmions (2π, 4π, 6π…), and skyrmion bags, analogous to those formed in magnetic systems. The feasibility of regular patterning, control, and read-out of weakly interacted skyrmions is revealed, and the maximum density is over 300 Gb/inch$^2$. Contrast to the relatively short lifetime of the skyrmion bubbles, the high-degree twisted skyrmions and the skyrmion bags with closed rings exhibit strong topology-enhanced stability. Experimental and theoretical analysis implies the existence of electric Dzyaloshinskii-Moriya interaction mediated by oxygen octahedral tiltings in the FM heterostructures, which therefore provide fertile playground for exploring exotic topological states and emergent phenomena. Our study sheds new insights into the polar topological textures in ferroelectrics and demonstrates the feasibility of developing robust storage and logic devices based on polar topological textures.

## 4. Experimental Section

*Sample Growth*: PTO/CCMO/SRO and PTO/SRO heterostructures were grown on single-crystal (001) STO substrates via PLD (KrF laser). The growth conditions for the bottom SRO layers and inserted CCMO layers were 650 °C and 680 °C under oxygen pressure of 12 Pa, respectively. The PTO layers were grown at 610 °C under oxygen pressure of 10 Pa. For all materials, the laser fluence was fixed at 1 J $cm^{-2}$ with a repetition rate of 5 Hz. After deposition, the heterostructures were cooled to room temperature at a rate of 10 °C $min^{-1}$ under oxygen pressure of 5 KPa.

*Structural and TEM Characterizations*: Structural characterization of the heterostructures was performed by XRD (PANalytical X'Pert PRO) with Cu-Kα radiation (wavelength λ = 1.5405 Å). Cross-sectional dark field TEM images and selected area electron diffraction patterns were acquired by using TEM (Titan Themis 300 S/TEM, FEI) operated at 300 kV with an objective aperture of 10 μm in diameter to select the transmitted beam. The high-resolution HAADF-STEM imaging provides spatial resolution adequate to measure the atomic positions of the A and B site cations of PTO. We used integrated differential phase contrast STEM (iDPC-STEM) to conduct atomic-column resolved observation of oxygen displacement. The *in-situ* experiments were carried out on the aberration-corrected FEI Tecnai F30 with an in situ STM-TEM holder (PicoFemto) at 300 kV in high-angle annular dark-field mode. An electrochemically etched tungsten tip serves as the top electrode of electrical switching, and the SRO/CCMO or SRO layers connected to the holder ground serve as the bottom electrodes. The current transients are recorded by a TELEDYNE LECROY WavePro HD high-precision oscilloscope (2.5GHz, 20GSa/s, 12-Bit) and micro-current amplifier (Femto, DLPCA-200).

*Scanning Probe Microscopy Measurement*: The surface topography of the samples was revealed by the AFM mode of Asylum Research MFP-3D SPM. The PFM measurements were carried out by using the same SPM system with Cr/Pt-coated conductive tips (Budget sensors). The PFM hysteresis loops were collected in the dual ac resonance tracking (DART) mode with triangle voltage pulse waveforms applied to the PFM tips. The current maps of the polar textures were characterized by CAFM via using conductive tips with a boron-doped diamond coating (Adama innovations). The tips have a typical curvature radius of 10 ± 5 nm, a nominal spring constant of 2.8 N/m, and a free-air resonance frequency of 65 kHz.

**Acknowledgements**

The authors sincerely thank Dr. Deyang Chen, Dr. Weiming Xiong and Dr. Congbing Tan for their helpful discussions. This work was supported by the National Natural Science Foundation of China (Nos. 12132020, 11972382, 12222214, 12072379 and 12302211), Guangdong Provincial Key Laboratory of Magnetoelectric Physics and Devices (No. 2022B1212010008) and the Shenzhen Science and Technology Program (Grant No. 202206193000001, 20220818181805001).

**Conflict of Interest**

The authors declare no conflict of interest.

**Data Availability Statement**

The data that support the findings of this study are available in the Supporting Information of this article.

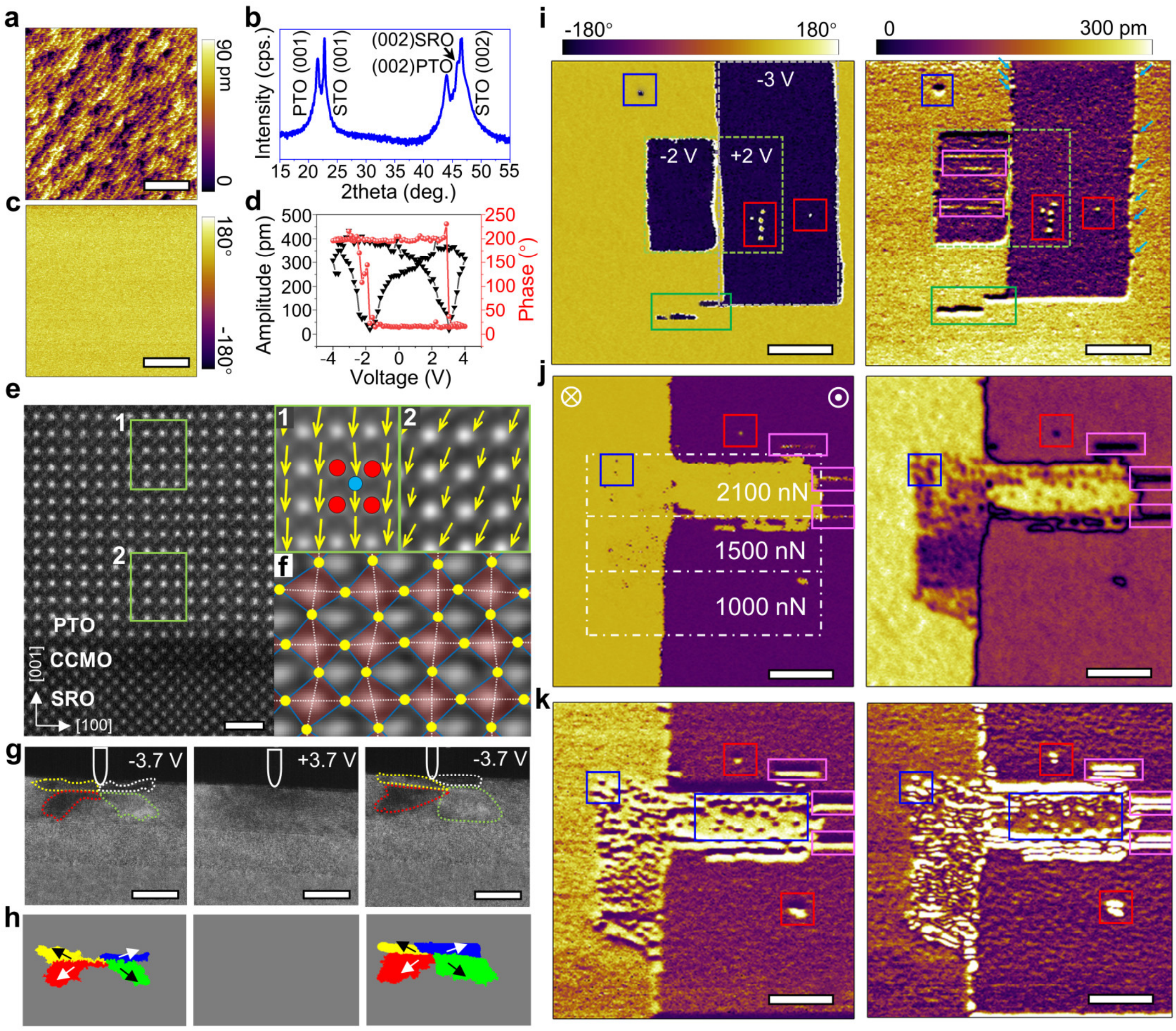

**Figure 1.** Structure of PTO/CCMO/SRO heterostructures and topological phase transition under electrical and mechanical stimulation exerted by PFM tip. a) AFM topography and b) XRD θ-2θ scan of PTO/CCMO/SRO thin films grown on (001)-oriented STO substrates. c) PFM phase image (scale bars, 1 μm) and d) PFM phase-voltage hysteresis loop (red curve) and butterfly-like amplitude-voltage loop (black curve) of the as-grown PTO thin film. e) HAADF-STEM images showing a homogeneous downward polarization state of the PTO thin film. Scale bar, 1 nm. f) iDPC-STEM image showing large distortion of the oxygen octahedra (as indicated by a blue rhombus) in the PTO thin film. g) Selected snapshots showing the creation of a center-divergent topological texture under ± 3.7 V pulsed bias exerted by a tungsten probe in an *in-situ* TEM experiment. Scale bars, 20 nm. h) Color mapping of the different contrast regions of the TEM images shown in (g). i) VPFM and LPFM phase images of the PTO film after applying a series of electrical stimulation (-3V, -2V and +2V) in selected regions, revealing topological phase transition induced by electric field. $Q = +1$ and $Q = -1$ Néel-type skyrmion-like bubbles emerge in the downward and upward domains as marked by blue and red boxes, respectively. Skyrmion-like bubbles and elongated skyrmion-like stripes also emerge at or near the domain walls (indicated by the blue arrows and pink/green boxes). Scale bars, 500 nm. j) VPFM phase, VPFM amplitude and k) LPFM phase, LPFM amplitude images of PTO film with both upward (written by -3 V external bias) and downward polarization (as-grown state) after scanning with different tip forces in a 1 μm × 1.5 μm area (white dashed box). $Q = \pm 1$ Néel-type skyrmion-like bubbles and elongated skyrmion-like stripes are marked by different color boxes, same as (i). Scale bars, 400 nm.

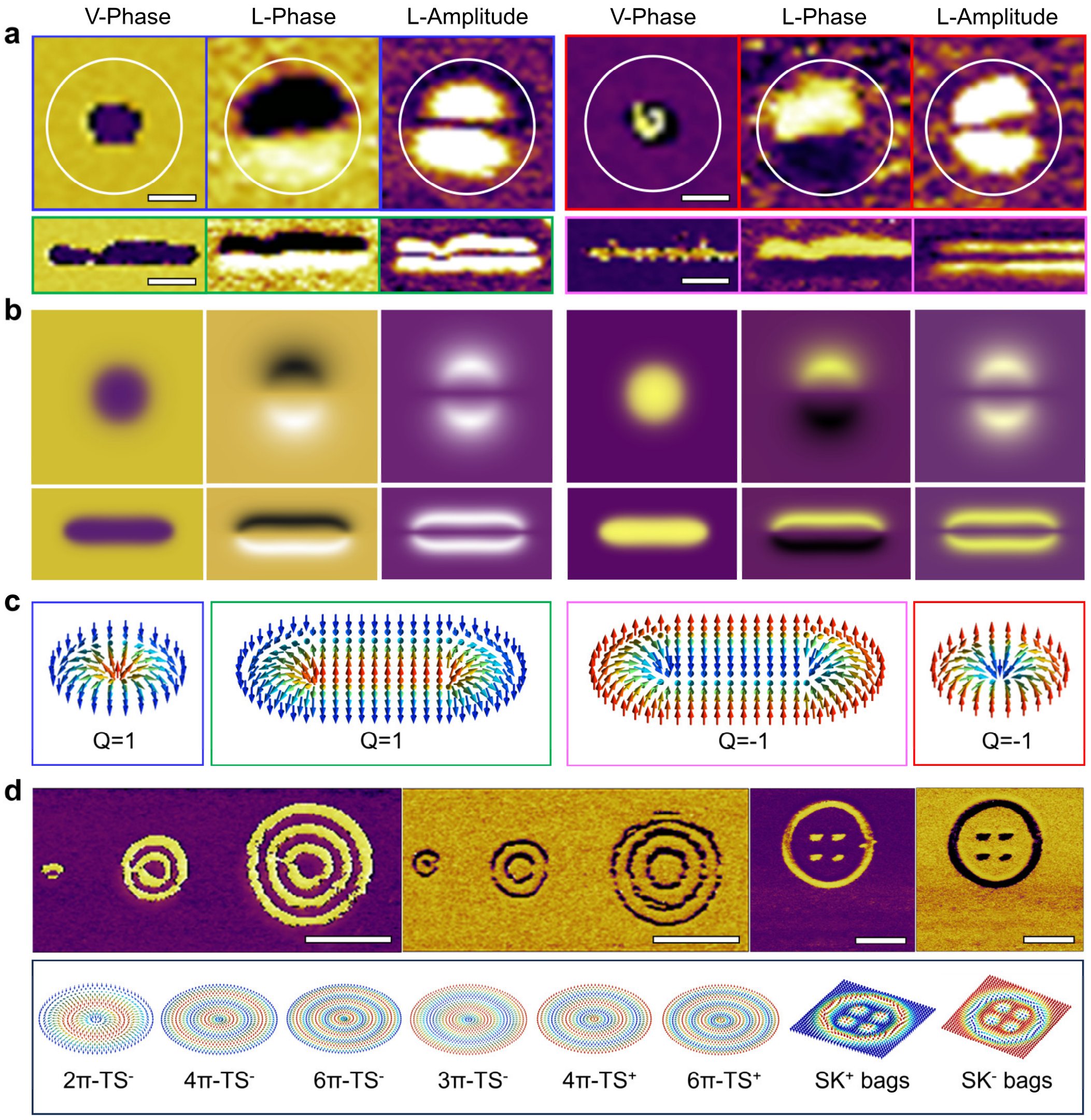

**Figure 2.** Diverse skyrmion-like textures constructed in PTO/CCMO/SRO heterostructures. a) Zoomed-in VPFM phase (V-phase), LPFM phase (L-phase) and amplitude (L-Amplitude) images of $Q = \pm 1$ skyrmion-like bubbles and elongated skyrmion-like stripes. Scale bars, 70 nm. b) Simulated PFM signals of the skyrmion-like textures in PTO thin film obtained by phase-field model. c) Schematics of $Q = \pm 1$ skyrmion-like bubbles and elongated skyrmion-like stripes. d) LPFM phase images of different types of high-degree skyrmion-like textures (denoted as $2\pi$-TS$^-$, $4\pi$-TS$^-$, $6\pi$-TS$^-$ and SK$^+$ bag) written in a downward domain background, and those written in an upward domain background (denoted as $\pi$-SK$^-$, $3\pi$-TS$^-$, $4\pi$-TS$^+$, $6\pi$-TS$^+$ and SK$^-$ bag) and corresponding configurations of the skyrmion-like textures (bottom panels). Scale bars, 600 nm.

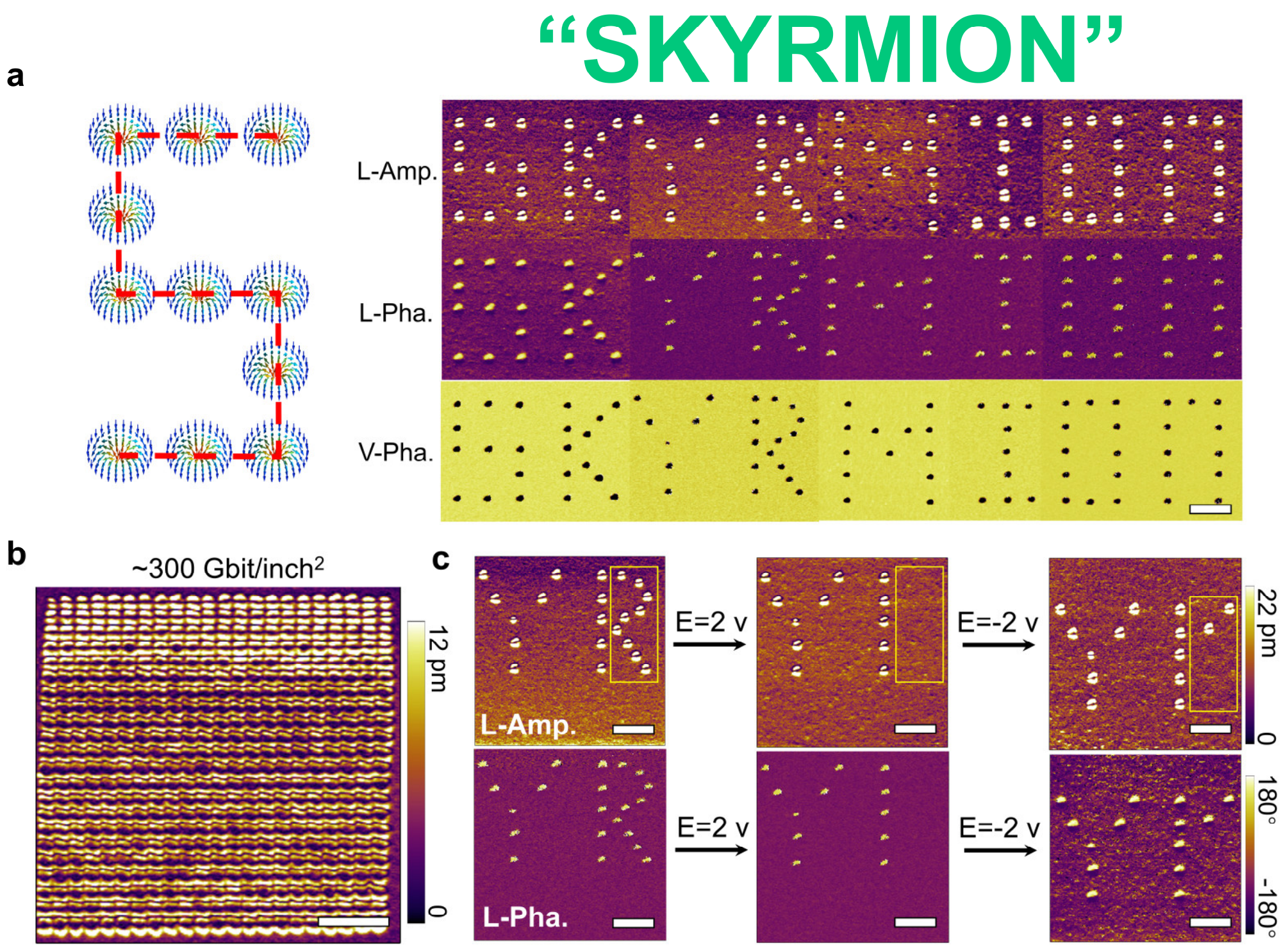

**Figure 3.** Regular patterning and control of skyrmion-like bubbles. a) Schematic of the capital letter “S” composed of 11 individual skyrmions and the LPFM amplitude, phase and VPFM phase images of word “SKYRMION”. b) The LPFM amplitude image of a regular pattern of skyrmions with a density ~300 Gbit/inch$^2$. c) The LPFM amplitude and phase images of letter “R”, showing the reversible erasure and writing of skyrmions. All the skyrmions were written by a single point loading of −2 V tip bias and erased by +2 V tip bias. Scale bars, 200 nm.

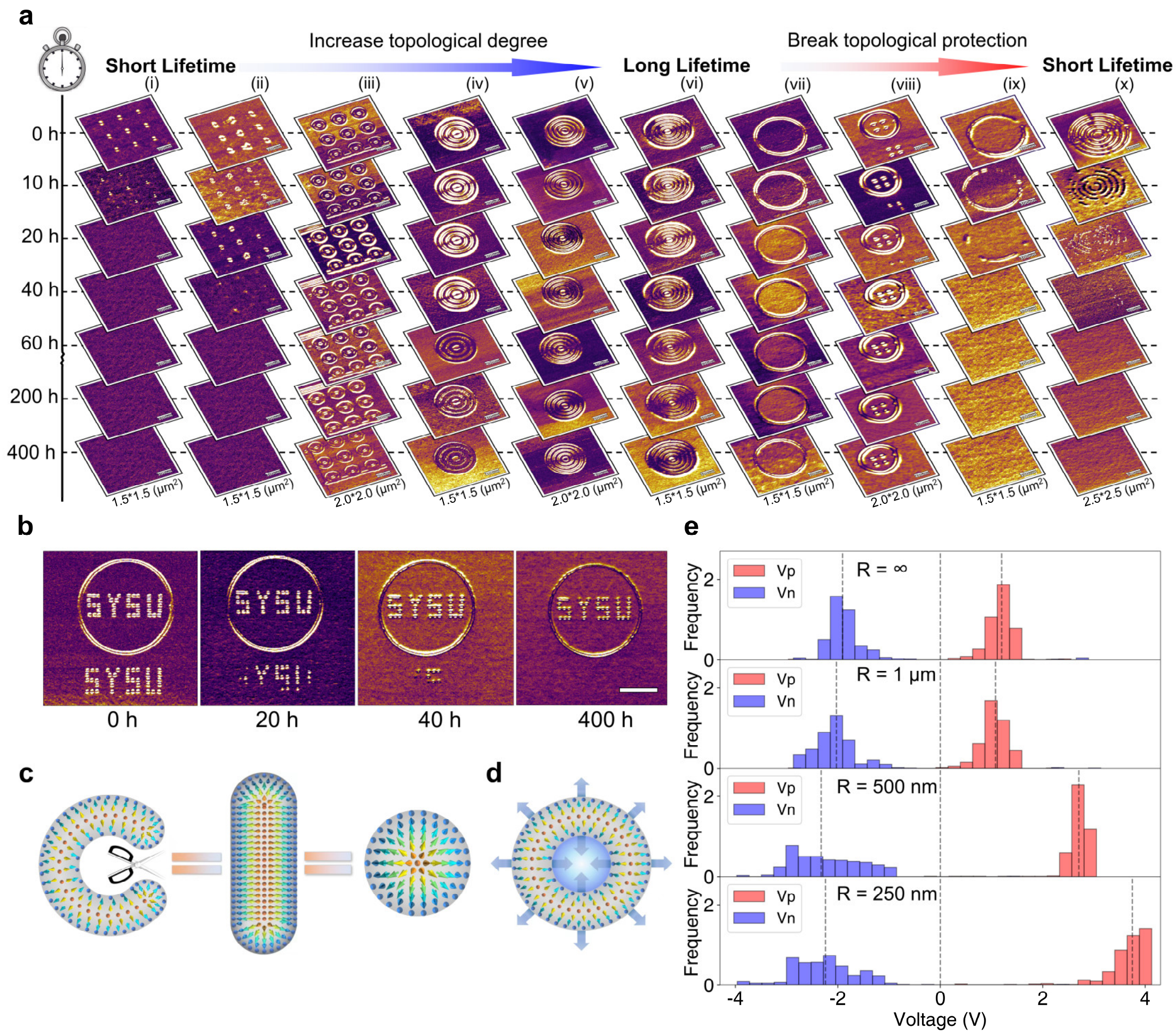

**Figure 4.** Topology-dependent stability of skyrmion-like textures created in PTO thin film. a) Temporal evolution of LPFM amplitude images of ten typical topological textures written in a downward domain background. b) Temporal evolution of LPFM amplitude image of π-SK$^+$ bag with one "SYSU" pattern enclosed by the ring and one outside the ring. Scale bar, 1 μm. c) Schematic illustration of the intrinsic topological protection of the ring structure. There should be a topological protective barrier between the closed ring and unclosed ring structures, and the latter is topologically equivalent to elongated skyrmion stripe or bubble. d) Schematic showing the appearance of a relatively compressive stress state inside the ring due to the electrostriction effect. e) The statistical coercive voltages inside the ring structure as a function of the ring radius. $V_n$ and $V_p$ refer to the negative and positive coercive voltages, respectively.

Supporting Information

# Creation of independently controllable and long lifetime polar skyrmion textures in ferroelectric-metallic heterostructures

*Fei Sun, Jianhua Ren, Hongfang Li, Yiwei Wu, Jianwei Liang, Hui Yang, Yi Zhang, Jianyi Liu, Linjie Liu, Mengjun Wu, Xiaoyue Zhang, Wenpeng Zhu, Weijin Chen,* and Yue Zheng**

## Supporting Information text

### Phase Field Simulation

Phase-field method was adopted to simulate skyrmion-like textures in PTO thin films. The spontaneous polarization field $\mathbf{P} = (P_1, P_2, P_3)$ was selected as the order parameter, the evolution of which was captured by the time-dependent Ginzburg-Landau (TDGL) equation $\partial P_i / \partial t = -L\delta F / \delta P_i$, where $L$ and $t$ are the kinetic coefficient and the time, respectively. The system' free energy is expressed as a functional of the order parameter $F = \int_V (f_{\mathrm{Land}} + f_{\mathrm{grad}} + f_{\mathrm{elec}} + f_{\mathrm{elas}} + f_{\mathrm{eDMI}})\mathrm{d}V + \int_S f_{\mathrm{surf}}\mathrm{d}S$, where $f_{\mathrm{Land}}$, $f_{\mathrm{grad}}$, $f_{\mathrm{elec}}$, $f_{\mathrm{elas}}$, $f_{\mathrm{eDMI}}$ and $f_{\mathrm{surf}}$ are the densities of Landau energy, gradient energy, electrostatic energy, elastic energy, electric DMI energy and surface energy, respectively. $f_{\mathrm{eDMI}}$ is expressed as $D\cdot(P_3(\nabla\cdot\mathbf{P})-(\mathbf{P}\cdot\nabla)P_3)$, with $D$ being the eDMI coefficient, the magnitude of which is taken to be 3 $\mathrm{Nm^2/C^2}$ in the simulations. The detailed expressions of the other energy densities and values of the parameters can be found in previous reports.[1-3] Three-dimensional (3D) mesh of $n_x\Delta\times n_y\Delta\times n_z\Delta$ was used, with a grid spacing of $\Delta = 0.4$ nm. Periodic boundary conditions were adopted along the in-plane directions. A misfit strain $u_m = -0.01$ was applied on the PTO film. The screening factor $\beta$ was set to be 0.9 for simulations of the electrostatic-field-driven formation of skyrmion-like textures,[4,5] whereas ideal short-circuited boundary condition ($\beta$ =1.0) was set for simulations of eDMI-driven formation of skyrmion-like textures. A single domain with a uniform out-of-plane polarization was used as the initial state. For the simulations of skyrmion-like textures, a nonzero built-in field within the interested region or a position-dependent electric potential at the PTO/electrode interfaces was considered, mimicking the effect of different electric conditions at the PTO/CCMO and PTO/SRO interfaces.

**DFT calculations**

To explore possible OOT-mediated eDMI in PTO thin films, a multi-layer model of PTO/SRO heterostructure was built (Figure S2). The interaction between Ti ions was believed to play a major role in eDMI, therefore we only calculated the eDMI vector between Ti ions.[6] A symmetrical interface configuration was chosen to maintain the inversion symmetry of the system without OOTs. Taking a similar method that was used to analyze magnetic DMI interactions,[7] we derived the eDMI interaction from the energy difference of two polarization configurations with opposite chiralities. Ground-state calculations of PTO/SRO heterostructures were performed with the VASP code utilizing a plane-wave basis set at an energy cutoff of 500 eV to ensure convergence.[8,9] The exchange-correlation was treated in the generalized gradient approximation with the Perdew-Burke-Ernzerhof parametrization (GGA-PBE).[10] Integrations over the first Brillouin zone were performed by discrete k-point sampling. A 3×4×3 Monkhorst–Pack mesh was used in this step, which was well-converged for the total energy calculations.[11] To calculate the eDMI interaction between Ti ions at the interface, we set up clockwise (CW) and anticlockwise (ACW) polarization configurations as shown in Figure S2, with the displacements of Ti ions being 0.1Å. In analog to magnetic DMI,[7] the in-plane eDMI vector between Ti ions can be written as $\boldsymbol{d}_{ij} = d \cdot (\hat{\boldsymbol{z}} \times \hat{\boldsymbol{u}}_{ij})$ with $\hat{\boldsymbol{z}}$ and $\hat{\boldsymbol{u}}_{ij}$ being unit vectors pointing along $z$-direction and from site $i$ to site $j$, respectively. Considering site equivalency, one can obtained $\Delta E_{\mathrm{DMI}} = E_{\mathrm{CW}} - E_{\mathrm{ACW}} = 32 d_{21}^{y} u^2$, where $E_{\mathrm{CW}}$ and $E_{\mathrm{ACW}}$ are the DFT total energies of heterostructures shown in Figure S2, and $d_{21}^{y}$ is the component of eDMI vector between Ti #1 and #2 ions in the $y$-direction.

**Supporting Information Figures and Table**

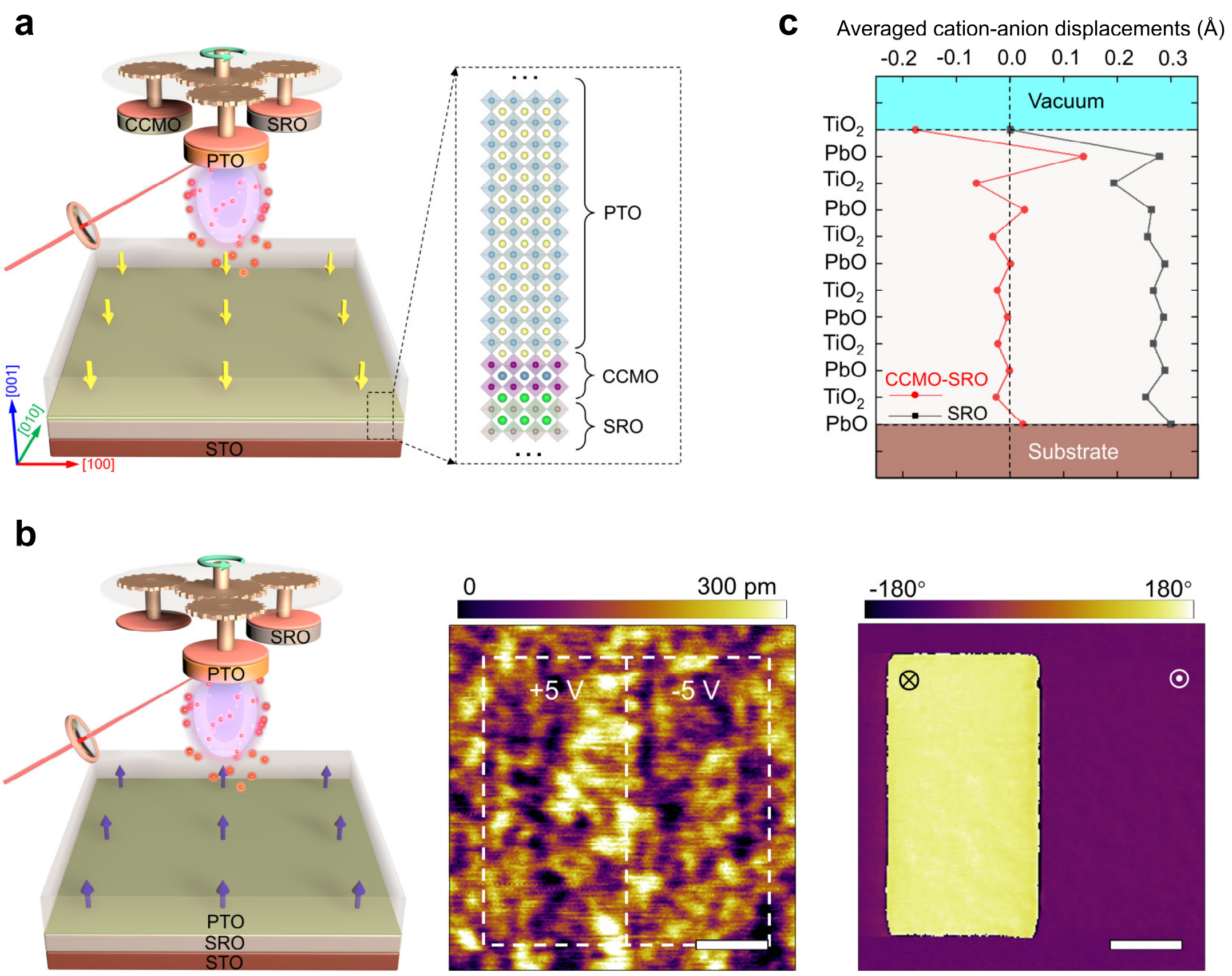

**Figure S1.** Structure diagrams of the PTO thin film before and after the insertion of CCMO layer. (a) Schematic of the PLD growth of PTO/CCMO/SRO heterostructures with a pristine downward polarization state. (b) (Left panel) Schematic of the PLD growth of PTO/SRO heterostructures with a pristine upward polarization state. (Right panels) The AFM and VPFM images of the PTO/SRO heterostructure after applying ±5 V tip voltage at the dash line regions. The pristine upward polarization state (purple-colored area) of the PTO film was fully switched to downward polarization (yellow-colored area) after applying a +5 V tip voltage. (c) The DFT calculated the averaged cation-anion displacements of the relaxed PTO/CCMO/SRO and PTO/SRO heterostructures, revealing the two heterostructures favor upward and downward polarization state, respectively. Scale bars in (b), 1 μm.

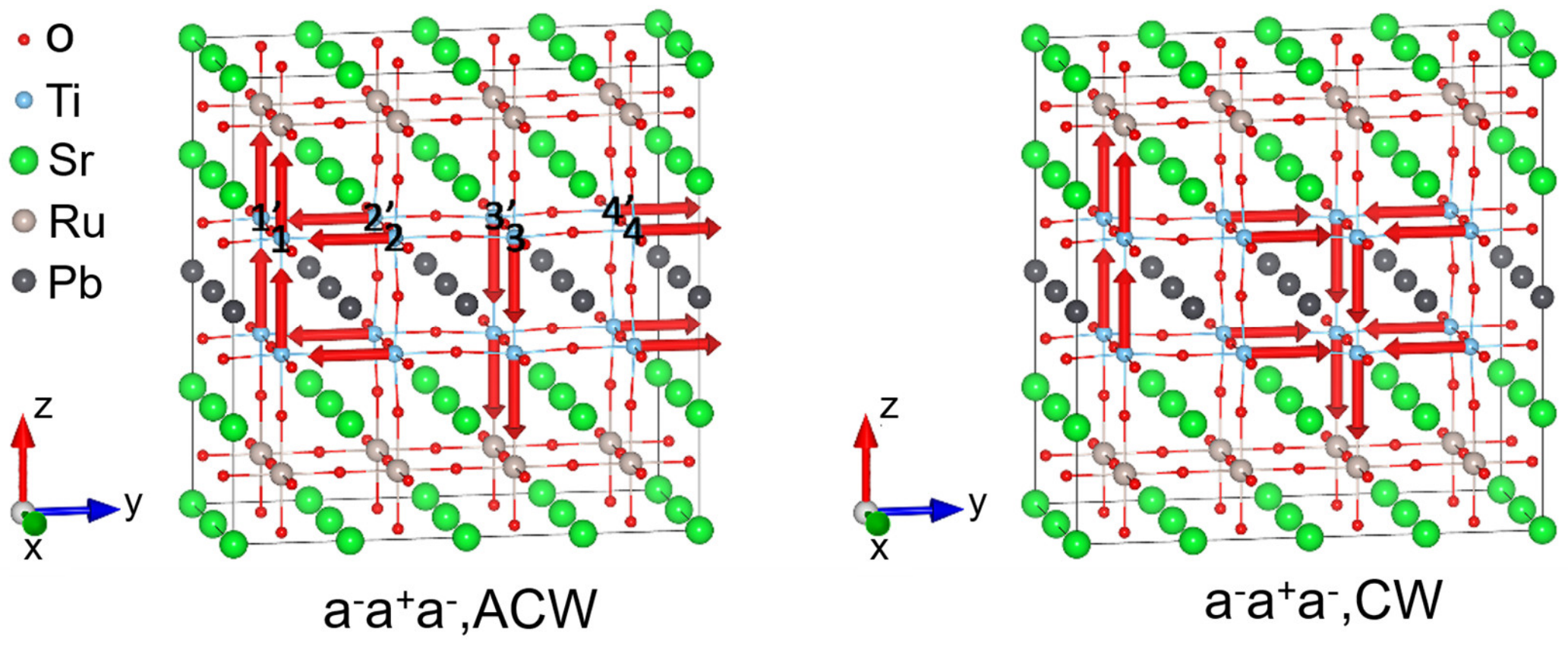

**Figure S2.** DFT calculation model for searching interlayer electric DMI mediated by $a^-a^+a^-$ mode oxygen octahedral tiltings (OOTs) in PTO/SRO heterostructure. Anticlockwise (ACW) and clockwise (CW) polarization configurations were used to derive eDMI interaction between interfacial Ti ions.

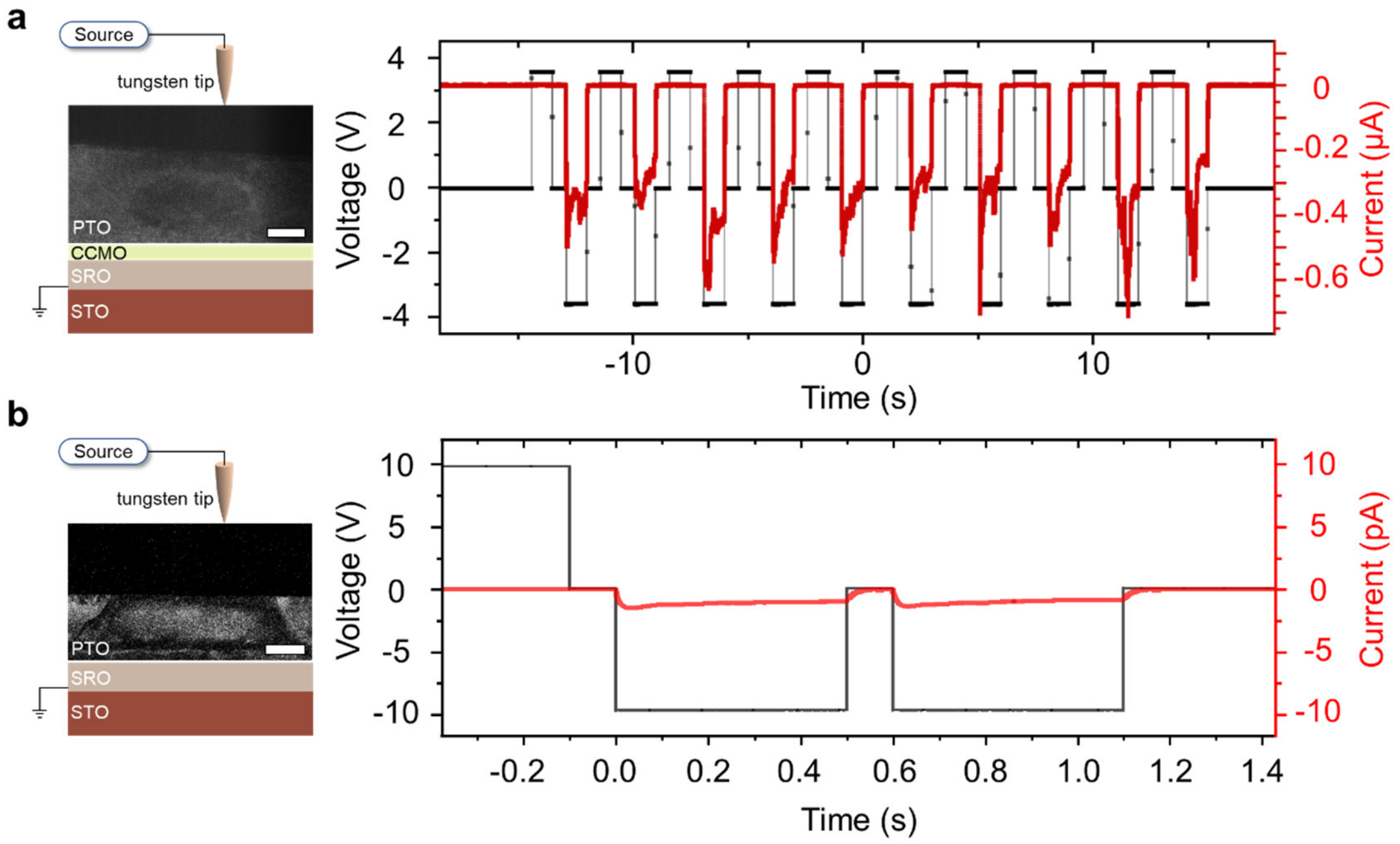

**Figure S3.** Domain switching and current evolution of two PTO thin film samples under local tip pulsed bias in *in-situ* TEM experiments. (a) TEM contrast of the switched polar texture (left panel) and current transients during respective domain switching (right panel) in the PTO/CCMO/SRO heterostructure. The evolution of the TEM contrast of this sample during the experiment is recorded in Movie S1, and typical snapshots are shown in Figure 1g. Scale bar, 10 nm. (b) Corresponding results of the PTO/SRO heterostructure. The evolution of the TEM contrast of this sample during the experiment is recorded in Movie S2. Scale bar, 200 nm.

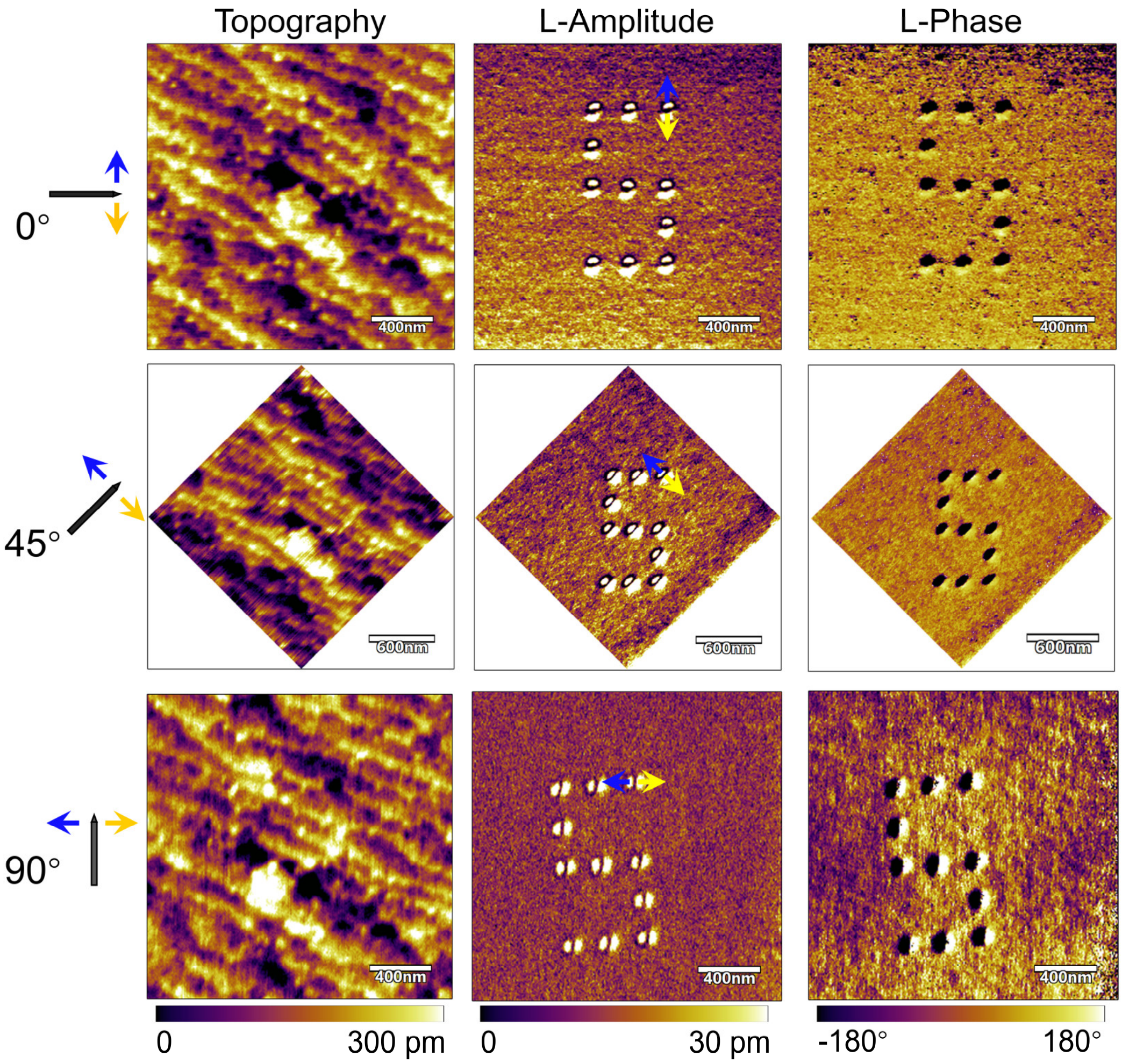

**Figure S4.** Reconstruction of the in-plane polarization distribution of a written pattern (letter "S") consisting of skyrmion-like bubbles in the PTO thin film by vector PFM. Topography, LPFM amplitude and LPFM phase images at the same region were measured with three different cantilever orientations ($\alpha = 0°$, $\alpha = 45°$, and $\alpha = 90°$, where $\alpha$ is the angle between the cantilever and [100] axis of the PTO film), showing the rotational invariance of the in-plane polarization distribution of the skyrmion-like bubbles.

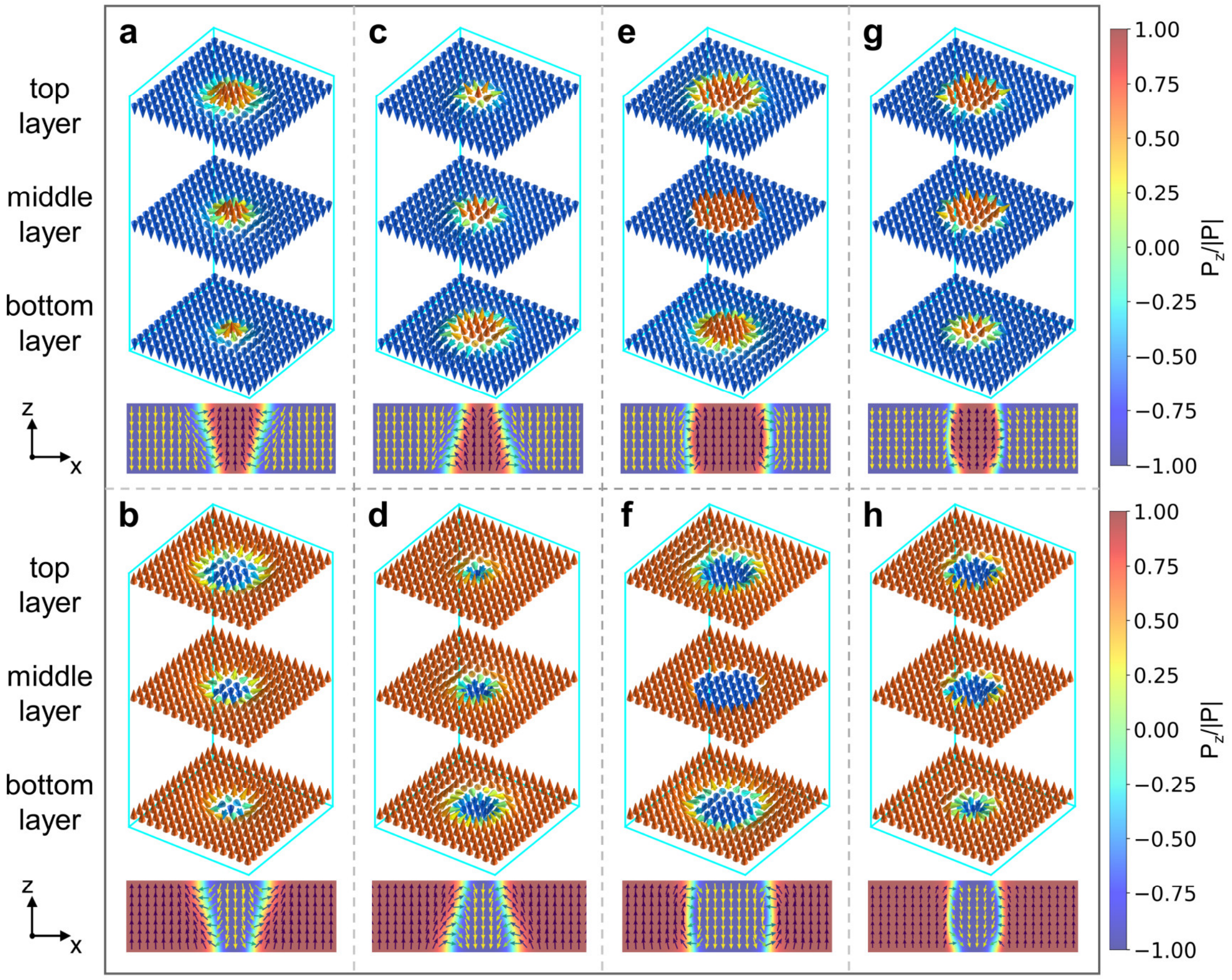

**Figure S5.** Phase-field simulations of the skyrmion-like bubble textures dominated by various mechanisms in PTO thin films. The top, middle and bottom layers of skyrmion-like bubbles, as well as the *x*-*z* cross-section configuration along the skyrmion center are presented. A single domain structure with a uniform out-of-plane polarization was used as the initial state. (a-d) Skyrmion-like bubbles driven by interface eDMI. (a) Positive eDMI and positive polarity. (b) Positive eDMI and negative polarity. (c) Negative eDMI and positive polarity. (d) Negative eDMI and negative polarity. (e-h) Skyrmion-like bubbles driven by electrostatic fields. A nonzero built-in field $E_b$ within the interested region (e and f) or a position-dependent electric potential $\varphi$ at the PTO/electrode interfaces (g and h) was considered.

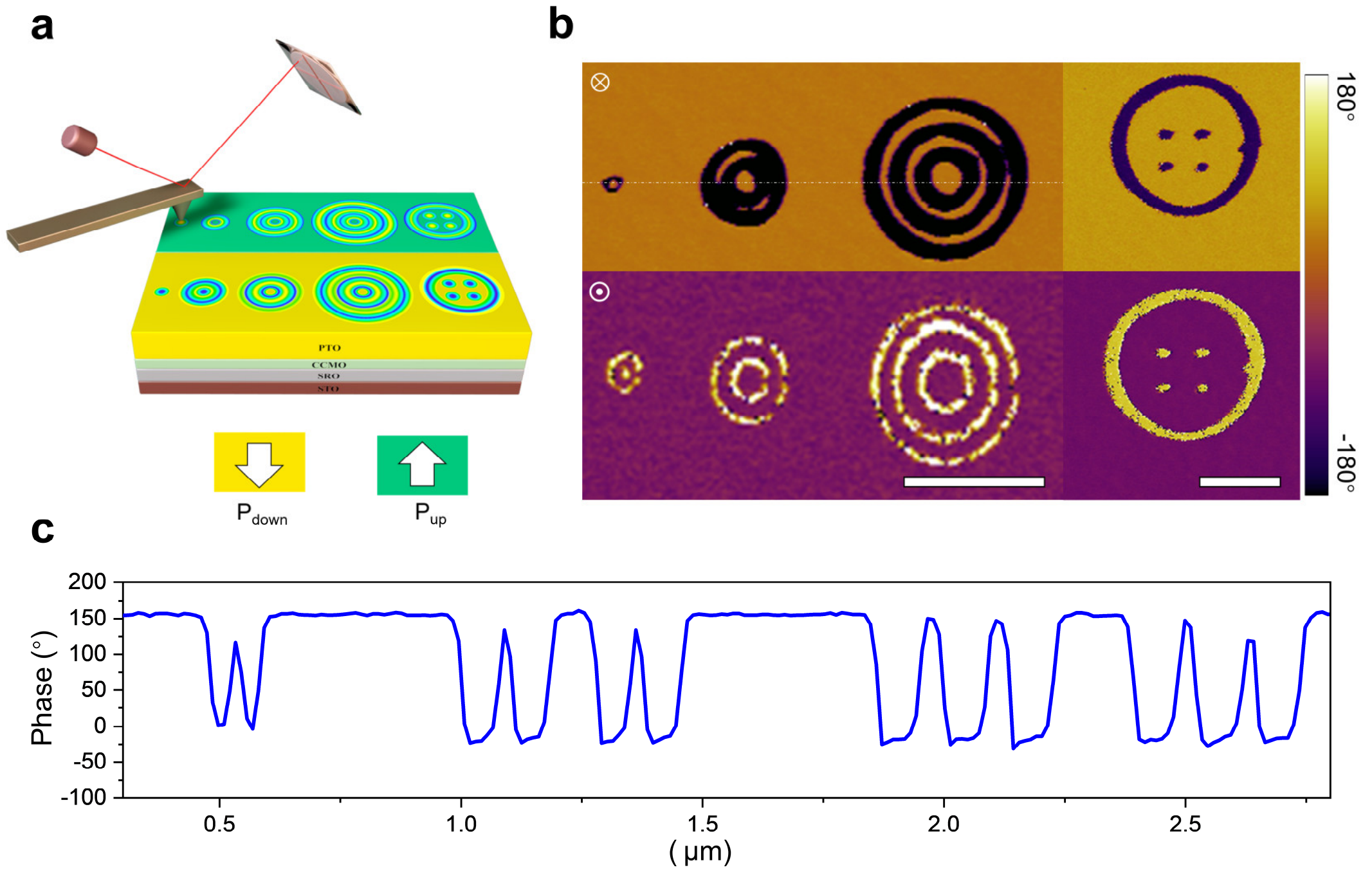

**Figure S6.** Diverse skyrmion-like textures constructed in PTO/CCMO/SRO heterostructures. (a) Schematic illustration of constructing different types of skyrmion-like textures in the PTO/CCMO/SRO heterostructure by local electrical and mechanical stimuli of a PFM tip. (b) VPFM phase images of different types of skyrmion-like textures (denoted as $2\pi$-TS$^-$, $4\pi$-TS$^-$, $6\pi$-TS$^-$ and SK$^+$ bag in Figure 2d) written in a downward domain background and those (denoted as $3\pi$-TS$^-$, $4\pi$-TS$^+$, $6\pi$-TS$^+$ and SK$^-$ bag in Figure 2d) written in an upward domain background. (c) Cross-section line profile of the VPFM phase image. Scale bars, 600 nm.

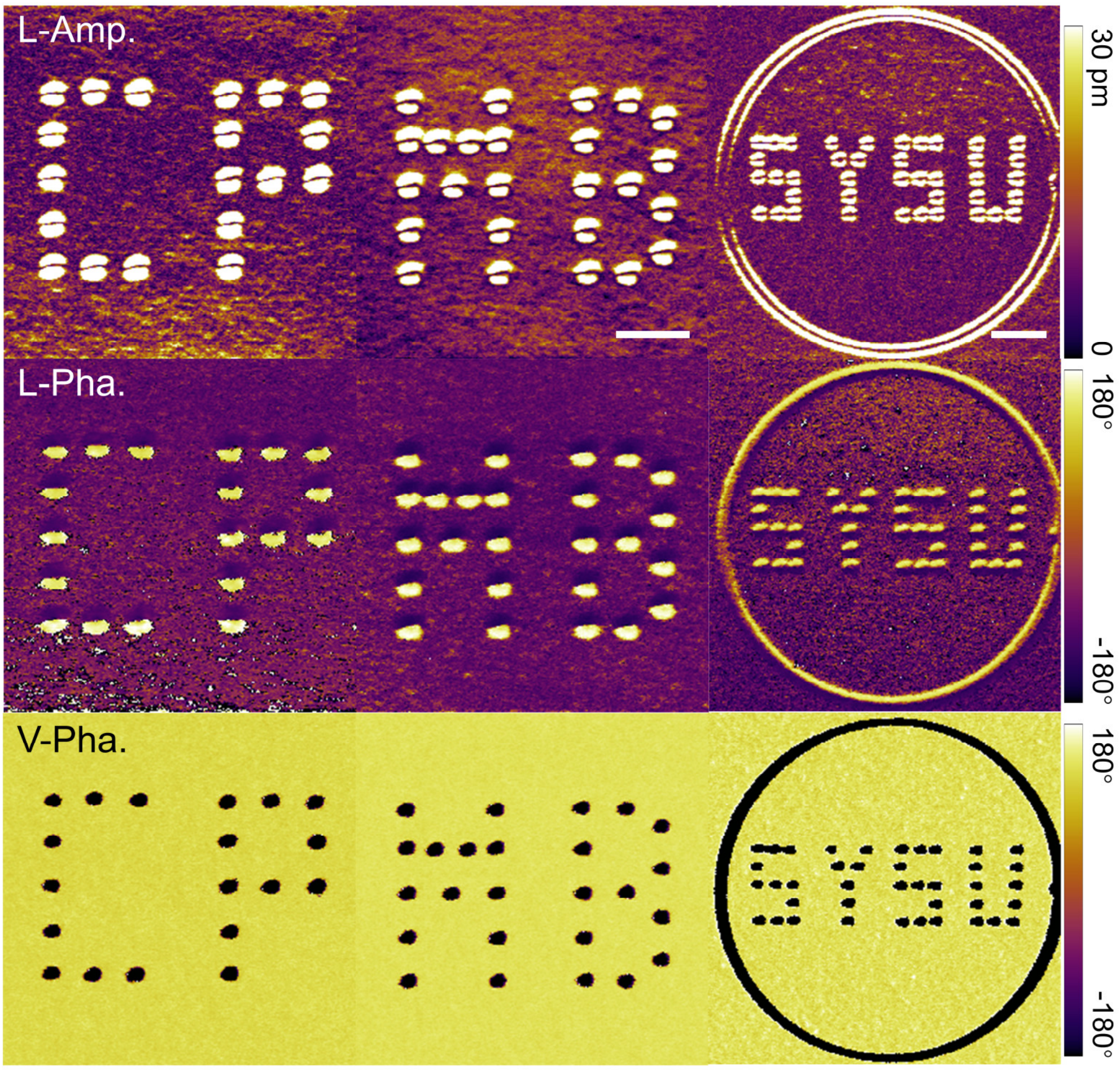

**Figure S7.** Deterministic control of polar skyrmion patterns. The LPFM amplitude, phase and VPFM phase images of two regular patterns of electrically written skyrmion-like bubbles. The word "CPMB" is the abbreviation of our laboratory name and "SYSU" is the abbreviation of our university name. Scale bars, 400 nm.

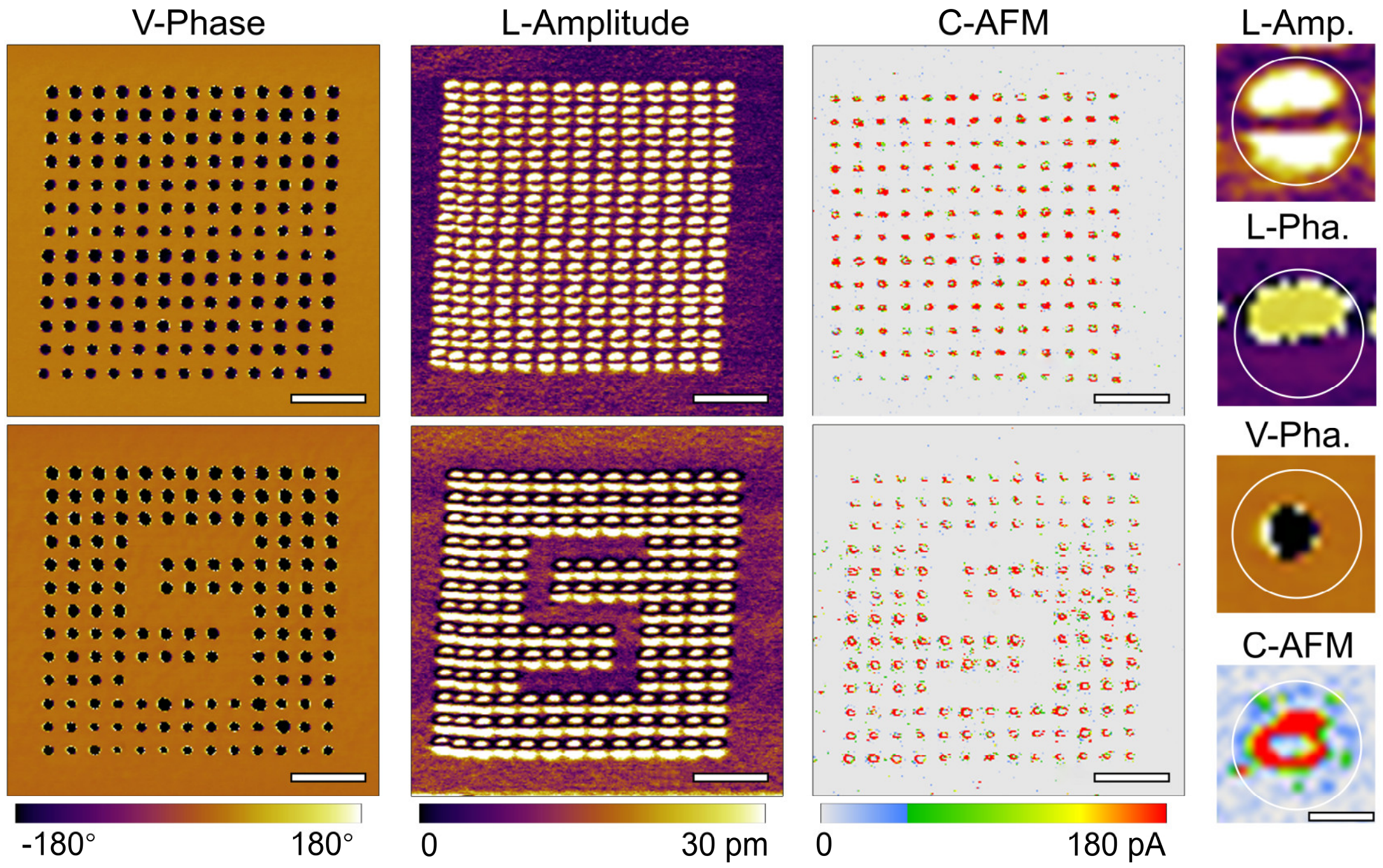

**Figure S8.** Transport characteristic of the written skyrmion patterns. The VPFM phase, LPFM amplitude and CAFM current maps (reading bias of -1.3V) were collected at the same area. By applying +2 V tip bias locally, some of the skyrmion bits were erased, as shown by the letter "S" with a low conductance. Scale bars, 200 nm (and 13 nm in zoom-in images).

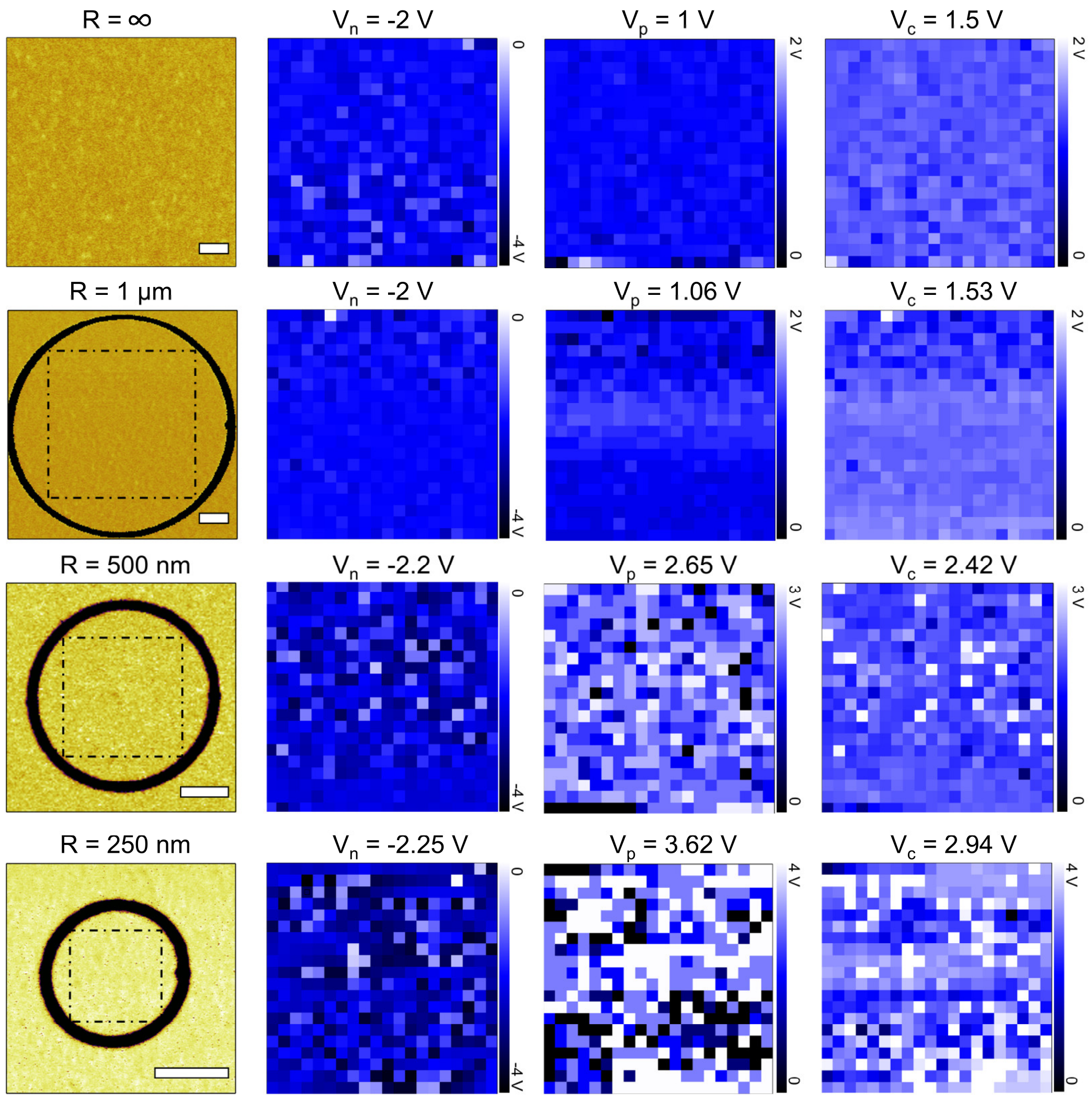

**Figure S9.** Phase contrast mappings of the PFM hysteresis loops collected at 20 × 20 grid positions of the inner region of polar ring structures with different radii. $V_c$ refers to the average coercive voltage, $V_n$ refers to the average negative coercive voltage and $V_p$ refers to the average positive coercive voltage. Scale bars, 250 nm.

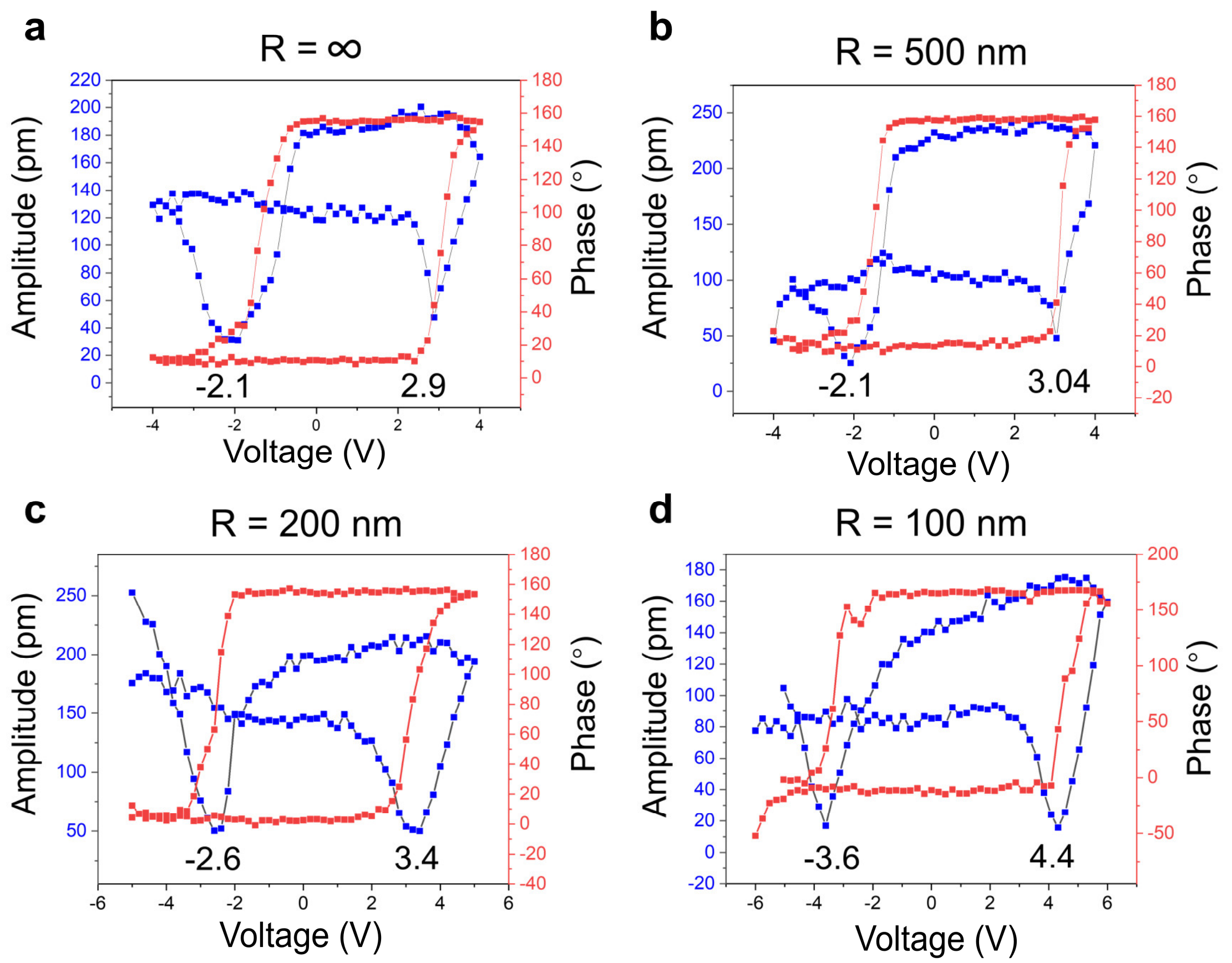

**Figure S10.** PFM hysteresis loops measured at the center of polar ring structures with different radii. (a-d) Thin film ($R = \infty$), $R$ = 500 nm, $R$ = 200 nm and $R$ = 100 nm.

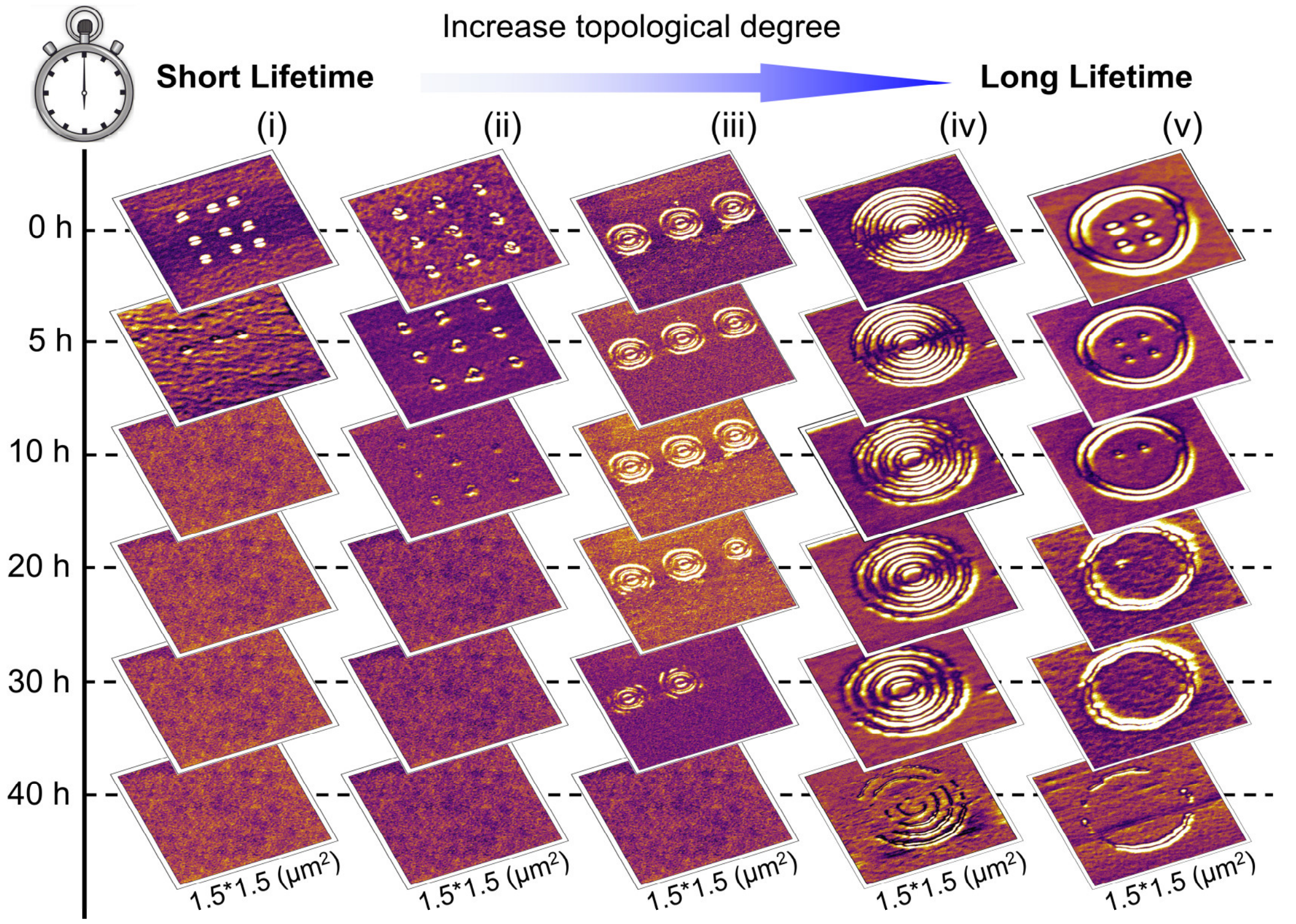

**Figure S11.** Temporal evolution of LPFM amplitude images of five topological textures written within a poled upward domain in the PTO thin film. (i) π-$SK^-$, (ii) 3π-$TS^-$, (iii) 4π-$TS^+$, (iv) 6π-$TS^+$, and (v) $SK^-$ bags.

**Table S1.** eDMI induced by various modes of oxygen octahedral tiltings (OOT). $d_{21}^{y}$ is the component of eDMI vector between Ti #1 and #2 ions in the *y*-direction as shown in Figure S2.

| **eDMI vector** | **Modes of OOT** | | | |
|---|---|---|---|---|
| | $a^0a^0a^0$ | $a^-a^-a^+$ | $a^+a^-a^-$ | $a^-a^+a^-$ |
| $\boldsymbol{d_{21}^{y}}$ **(eV/Å$^2$)** | 0 | $5\times10^{-7}$ | $-4.34\times10^{-8}$ | -0.0037 |

**Movie S1.**

Local electrical switching dynamics of the PTO/CCMO/SRO heterostructure in an *in-situ* TEM experiment.

**Movie S2.**

Local electrical switching dynamics of the PTO/SRO heterostructure in an *in-situ* TEM experiment.

## Supporting Information References